\documentclass[prb,aps,amssymb,showpacs,twocolumn,amssymb,amsmath]{revtex4}
\usepackage{graphicx}
\usepackage{dcolumn}
\usepackage{bm}
\usepackage{psfrag}

\newcommand{\be}{\begin{equation}}
\newcommand{\ee}{\end{equation}}
\newcommand{\bea}{\begin{eqnarray}}
\newcommand{\eea}{\end{eqnarray}}

\newcommand{\s}{\sigma}

\newcommand{\ba}{\begin{array}}
\newcommand{\ea}{\end{array}}
\def\nn{\nonumber\\}

\begin{document}

\title{Sign changes and resonance of intrinsic spin Hall effect
in two-dimensional hole gas}

\author {Tianxing Ma$^{1}$\footnote{txma@fudan.edu.cn}, Qin Liu$^{1,2}$}
\affiliation{$^{1}$Department of Physics, Fudan University, Shanghai
200433, China\\
 $^{2}$Department of Physics, CCNU, Wuhan 430079, China}
\date{\today}

\begin{abstract}
The intrinsic spin Hall conductance shows rich sign changes by
applying a perpendicular magnetic field in a two-dimensional hole
gas. Especially, a notable sign changes can be achieved by adjusting
the characteristic length of the Rashba coupling and hole density at
moderate magnetic fields. This sign issue may be easily realized in
experiments. The oscillations of the intrinsic spin Hall conductance
as a function of 1/$B$ is nothing else but Shubnikov-de Haas
oscillations, and the additional beatings can be quantitatively
related to the value of the spin-orbit coupling parameter. The
Zeeman splitting is too small to introduce effective degeneracy
between different Landau levels in a two-dimensional hole gas, and
the resonant intrinsic spin Hall conductance appears in high hole
density and strong magnetic field due to the transition between
mostly spin-$-\frac{1}{2}$ holes and spin-$\frac{3}{2}$ holes is
confirmed. Two likely ways to establish intrinsic spin Hall effect
in experiments and a possible application are suggested.
\end{abstract}

 \keywords{}

\maketitle
In the developing field of spintronics\cite{Prinz98Science}, which
is believed to be a promising candidate for future information
technology\cite{Review}, the generation and efficient control of
spin current is a key issue. So the intrinsic spin Hall effect
(ISHE) predicted by Murakami {\it et al.}\cite{Murakami03Science}
and Sinova {\it et al.}\cite{Sinova04} has generated intensive
theoretical studies and this ISHE is associated with the
dc-field-induced transitions between the spin-orbit-coupled (SOC)
bands, and contributes from all occupied electron states below the
Fermi energy. The ISHE is distinguished from the extrinsic spin Hall
effect (ESHE), which is due to impurity scattering\cite{Hirsch}. On
the experimental side, the spin Hall effect (SHE) reported in a
two-dimensional hole gas (2DHG) is likely of the intrinsic
origin\cite{Wunderlich,Bernevig}; nevertheless, the intrinsic or the
extrinsic origin of the SHE observed in a 3D electron
film\cite{Kato} is still under dabate\cite{Engel,Bernevig0}.

The intrinsic spin Hall conductance (ISHC) of a 2DHG in a
perpendicular magnetic field has been studied within $k$-cubic
Rashba model by M. Zarea {\it et al.}\cite{Zarea} and a generalized
Luttinger model by us\cite{Ma} most recently. The work by M. Zarea
{\it et al.} only focused on the low field regime, and a
remark\cite{Sch} has been given by J. Schliemann on this aspect: in
any case, influence of a magnetic field coupling to the orbital
degrees of freedom should only be appreciable if the field is strong
enough to produce typical cyclotron radii being of order of the
system size or smaller; therefore, arbitrarily small fields cannot
be expected to have an effect in real experiments. Within a more
general band structure Hamiltonian, we have predicted that the
``effective" energy crossing near Fermi energy between mostly
spin-$-\frac{1}{2}$ holes and spin-$\frac{3}{2}$ holes at a typical
magnetic field gave rise to a resonant ISHC\cite{Ma} with a certain
hole density, however, a strong magnetic field region is required.
Except resonance\cite{Ma,Shen,Dai}, the sign issue is another
crucial aspect of ISHC\cite{Yao,Shens}, which can be modulating by
external magnetic field\cite{Ma}. To be convenient for future
experimental detection or possible applications, it is of course
desirable to discuss the effect of magnetic field in detail,
especially in moderate magnetic field region, which can be easily
achieved in experiments or technologies.

In this paper, we will concentrate on the sensitivity of ISHC to
magnetic field, the length scale of the Rashba coupling, and hole
density, which shows rich sign changes. We predict that a more
notable sign changes can be achieved by adjusting the length scale
of the Rashba coupling at certain moderate magnetic fields for a
fixed hole density. By discussing the effect of Zeeman splitting on
energy levels, we clarify that the interplay between mostly
spin-$-\frac{1}{2}$ holes and spin-$\frac{3}{2}$ holes is the origin
of resonant ISHC in a 2DHG under a perpendicular magnetic field,
which is different from the case\cite{Shen} in two-dimensional
electron gas (2DEG). In addition, the oscillations of the intrinsic
spin Hall conductance as a function of 1/$B$ is nothing else but
Shubnikov-de Haas (SdH) oscillations, and the additional beatings
can be quantitatively related to the value of the SOC parameter. Two
likely ways to establish ISHE in experiments and a possible
application are suggested, which may be easily realized in
experiments.

When the quasi 2DHG is sufficiently narrow and the density and the
temperature are not too high, only the lowest heavy-hole subband is
occupied. The effective Hamiltonian for a single heavy hole
subjected a spin-orbit interaction due to structural-inversion
asymmetry (SIA) can be written as\cite{Winkler0,Loss,Liu}
\begin{equation}
H_{0}=\frac{\vec p^{2}}{2m}+i\frac{\alpha}{\hbar^{3}}
\left(p_{-}^{3}\sigma_{+}-p_{+}^{3}\sigma_{-}\right)\,
\label{defham},
\end{equation}
using the notations $p_{\pm}=p_{x}\pm ip_{y}$,
$\sigma_{\pm}=(\sigma_{x}\pm i\sigma_{y})/2$, where $\vec p$, $\s$
denote the hole momentum operator and Pauli matrices respectively.
These Pauli matrices operate on the total angular momentum states
with spin projection $\pm 3/2$ along the growth direction, and we
simply call it holes with spin-$\pm\frac{3}{2}$ below. In the above
equation, $m$ is the heavy hole mass, and $\alpha$ is Rashba SOC
coefficient due to SIA. The validity of above
model\cite{Winkler0,Loss,Liu} given by the Hamiltonian (1) was
restricted to sufficiently small wave numbers and densities.

We follow the method by M. Zarea\cite{Zarea}{\it et al.} and
us\cite{Ma} bellow. We impose a magnetic field $\vec{B}=B\hat{z}$ by
choosing the Landau gauge $\vec{A}=-yB\hat{x}$, and then $p_{x}=
\hbar k_{x}+\frac{eB}{c}y$, where $-e$ is the electric charge. The
destruction operator
$a=\frac{1}{\sqrt{2m\hbar\omega}}(p_{x}+ip_{y})$, and the creation
operator $a^{\dag}=\frac{1}{\sqrt{2m\hbar\omega}}(p_{x}-ip_{y})$ are
introduced to describe Landau levels, where $\omega=\frac{eB}{mc}$.
These operators have the commutation $[a,a^{\dag }]=1$. In terms of
these operators, the Hamiltonian in the absence of Zeeman coupling
is
\begin{equation}
H_{0}=\hbar\omega[(a^{\dag}a+\frac{1}{2})
+i\gamma(a^3\s^+-{a^{\dag}}^3\s^-)],
\end{equation}
in which the dimensionless parameter $\gamma$ is defined by
$\gamma=2\frac{\alpha m}{\hbar^{2}}\sqrt{\frac{2eB}{\hbar c}}$. In
units of $\hbar\omega$, the eigensystem can be expressed as \bea
&&\psi_{ns}=\left(\ba{c}-i\cos\theta_{ns} \phi_{n-3} \\
\sin\theta_{ns}\phi_{n}\ea \right),\nn &&E_{ns}=(n-1)+s\sqrt{\gamma
^{2}n(n-1)(n-2)+\frac{9}{4}},\label{states} \eea with $s=\pm 1$ for
$n\geq 3$, $s=1$ for $n<3$ and $\phi_{n}$ is the eigenstate of the
$n$th Landau level in the absence of SOC. Besides,
$\theta_{ns}=\arctan[\frac{2\gamma
\sqrt{n(n-1)(n-2)}}{s\sqrt{4\gamma^{2}n(n-1)(n-2)+9}+3}]$, where
$\theta_{ns}\in(0,\frac{\pi}{2})$ for $n>3$, and
$\theta=\frac{\pi}{2}$ when $n<3$.

When the system is driven by a dc electric field in $-\vec{y}$
direction, the most interesting case is the $z$-direction-polarized
spin current along $x$ direction, and the corresponding spin current
operator can be defined\cite{Liu,Loss} as
$j_{S}=\frac{3}{2}\hbar\{v_x,\s^z\}$ where $v_{x}$ is the velocity
operator. We shall only focus on the first order in the perturbation
term because $<n,s|j_{S}|n,s>=0$, which is
\begin{eqnarray}
(j_{S})_{ns}=\sum_{n^{\prime }s^{\prime }}\frac{\left\langle
n,s\right\vert j_{S}\left\vert n^{\prime },s^{\prime }\right\rangle
\left\langle n^{\prime },s^{\prime }\right\vert H^{\prime
}\left\vert n,s\right\rangle}{\epsilon _{ns}-\epsilon_{n^{\prime
}s^{\prime }}}+H.c.,
\end{eqnarray}
where $(j_{S})_{ns}$ is the current carried by a hole in the state
$|n,s>$ of $H_{0}$, and $\epsilon_{ns}=\hbar\omega E_{ns}$. In
addition, $n'=n\pm 1$ since the matrix element vanishes for other
values of $n'$. The average current density of the $N_{h}$-hole
system is given by
$I_{S}=\frac{1}{L_{y}}\sum_{ns}(j_{S})_{ns}f(\epsilon _{ns})$ where
$f$ is the Fermi distribution function, and $N_{h}=\sum_{ns}f(
\epsilon _{ns})$. The ISHC is then given by $G_{S}=I_{S}/EL_{x}$,
where the Landau degeneracy factor $eB/(hc)$ is also included. We
can finally obtain the complete ISHC, and the 2DHG system we
consider is confined in the $x-y$ plane of an area $L_{x}\times
L_{y}$.
\begin{figure}
\begin{center}
\includegraphics[scale=0.45]{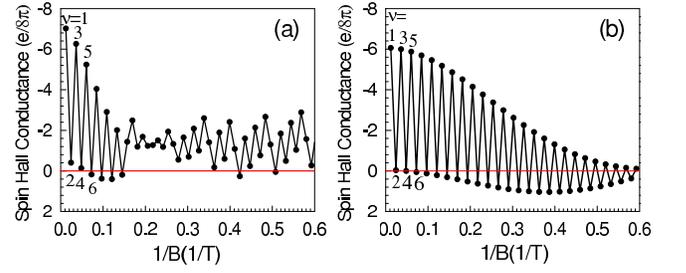}
\caption{(a) The ISHC as a function of $1/B$ for $n_{h}=2\times
10^{16}$\,m$^{-2}$, and the characteristic length $\alpha
m/\hbar^{2}=0.25$nm of the Rashba coupling\cite{Liu}. (b) Same as
Fig1.(a) but $\alpha m/\hbar^{2}=0.075$nm. }\label{chargedensity}
\end{center}
\end{figure}

Experimental study\cite{Wunderlich} has been performed for
heavy-hole density $n_{h}=2\times 10^{16}$\,m$^{-2}$, and the
characteristic length $\alpha m/\hbar^{2}=0.25$nm of the Rashba
coupling\cite{Liu}. By using these parameters, the ISHC as a
function of $1/B$ is shown in Fig.1 (a). The ISHC oscillates as a
result of the alternative occupation of mostly spin-$\frac{3}{2}$
holes and mostly spin-$-\frac{3}{2}$ holes, which is consistent with
our previous result\cite{Ma}. The oscillations of the ISHC as a
function of $1/B$ are nothing else but another manifestation of SdH
oscillations. Following the same argument of usual
SdH\cite{Winkler}, the periodicity of density of states with $1/B$
results from the equality of the area of two successive electron
orbits in $\vec{k}$-space and the oscillation frequency is given by
$f$=$\frac{n_h \phi_0}{2}$=41.36, where $\phi_0$=$\frac{h}{e}$ is
the flux quantum. In relative weak magnetic field region
($1/B>0.22/T$ in Fig.1 (a)), there are two independent oscillation
frequencies originating from the two spin-splitting subbands,
$n_h^\pm$=$\frac{1}{\phi_0}f_\pm$=$\frac{1}{4\pi}(k^{\pm}_f)^{2}$,
where $n_h^+$($n_h^-$) are respectively the hole density in spin up
(down) subbands and $k_f^\pm$ are respectively their Fermi momenta.
It is just the composition of these two frequencies produces the
additional beating pattern appearing in our numerical result, and
the additional beatings can be quantitatively related to the value
of the SOC parameter\cite{Loss}
\begin{eqnarray}
k_f^{\pm}&=& \mp\frac{1}{4L}(1-\sqrt{1-16\pi n_h L^2}) \nn
&+&[-\frac{1}{8L^2}(1-\sqrt{1-16\pi n_h L^2})+3\pi n_h]^{1/2},
\end{eqnarray}
where $L=\alpha m/\hbar^{2}$. Fast Fourier transformation has been
carried out separately on the strong- and weak magnetic field data
shown in Fig.1 and frequencies obtained agree numerically with Eq.
(5) and $f_\pm$=$\frac{h}{4\pi e}(k^{\pm}_f)^{2}$ rather well.

The importance of sign changes of ISHC has been emphasize by Yao
{\it et al.}\cite{Yao} by applying first-principles calculations to
study SHE in semiconductors and simple metals in the absence of a
magnetic field. For the definition of sign, positive spin Hall
conductance means that spin-up component flows to the positive $x$
direction. The sign of ESHE induced in the skew scattering, which
dominates over the side jump process in the weak disorder limit,
depends on the sign of the impurity potential, and does not change
with the carrier density or the Rashba coupling strength. As shown
in Fig.1(a), the sign of ISHC changes periodically between $\nu=6$
and $\nu=13$. Furthermore, this sign changes is robust in this case,
and a more notable sign changes is sensitive to $\alpha m/\hbar^{2}$
for a fixed hole density, which can be achieved by modulating
$\alpha m/\hbar^{2}$ at certain magnetic fields. Fig.1 (b) accounts
for this. A more notable sign changes is shown for the same hole
density as in Fig.1(a) at $\alpha m/\hbar^{2}=0.075$nm, where the
magnetic field is in the range $2.07T\sim 3.60T$, especially when
$2.50T<B< 2.85T$.

To be convenient for future experimental detection, we have made an
scan on the magnetic field, and the characteristic length scale
dependence on the most notable sign changes as a function of hole
densities, and the optimal $\alpha m/\hbar^{2}$ for the most notable
sign changes as a function of a hole density has been shown in
Fig.2, beside, the corresponding most favorable range of magnetic
field is also plotted as a function of hole densities, which denoted
by shadow in Fig.2. Since the ISHC in $+x$ direction is larger than
the ISHC in $-x$ direction for most case, the optimal $\alpha
m/\hbar^{2}$ and the corresponding most favorable range of $B$ for
the most notable sign changes means that when the ISHC in $-x$
direction is the largest for a fixed hole density.
\begin{figure}
\begin{center}
\includegraphics[scale=0.4]{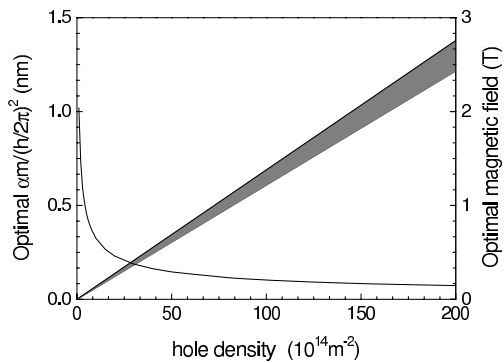}
\caption{The optimal $\alpha m/\hbar^{2}$ for the most notable sign
changes as a function of hole densities (solid line), and the
corresponding most favorable range of magnetic field is also plotted
as a function of hole densities (shadow).}\label{chargedensity}
\end{center}
\end{figure}

Moreover, The calculated ISHC shows sign changes as hole density
varies for a fixed magnetic field and the characteristic length
scale of Rashba coupling, which differs from the ESHE too. As shown
in Fig.3(a), the importance of this aspect is that, the required
magnetic field ($B=0.827T$) is easy to achieve in experiments, and
the sign changes of ISHC may be detected expediently. It's a likely
way to establish the ISHE, and it also means a possible application
in the future. If we can take the direction of the spin Hall current
as a new sign, i.e, $+x$ direction means ``0" and $-x$ direction
means ``1", and these two states maybe the basic for a new logic
electronic-device\cite{Ma}.

\begin{figure}
\begin{center}
\includegraphics[scale=0.45]{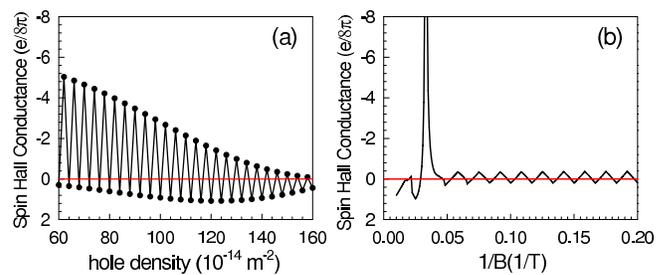}
\caption{(a) The ISHC as a function of hole densities for $B=0.82T$
and $\alpha m/\hbar^{2}$=0.05nm. (b) Resonant ISHC within Luttinger
model including Zeeman splitting, and parameter are taken from
Ref.\cite{Bernevig,Ma} except $n_{h}=2.9\times 10^{16}$\,m$^{-2}$,
and $g_{s}=-0.44$. }\label{chargedensity}
\end{center}
\end{figure}
The resonant ISHC predicted by us\cite{Ma} is caused by the energy
crossing near the Fermi energy due to the transition between mostly
spin-$-\frac{1}{2}$ holes and spin-$\frac{3}{2}$ holes. However, for
the 2D heavy-hole system described by a $k$-cubic Rashba model, we
can deduce from Eq. (3) that
\begin{eqnarray}
E_{ns}-E_{n\pm 1,\mp s}\neq0
\end{eqnarray}
is steady for any $n$ and Rashba coupling constant in the absence of
Zeeman splitting. Then there is NO resonant ISHC within Hamiltonian
(2) in the presence of a dc field.

Besides, the resonant ISHC was also predicted in 2D quantum wells in
a strong perpendicular magnetic field, where the Zeeman splitting
plays a crucial role. The $k$-linear Rashba coupling, generated by
spin-orbit interaction in wells lacking bulk inversion symmetry,
competes with Zeeman splitting to introduce additional degeneracies
between different Landau levels ($i$.$e$., two states $E_{n \mu}$,
$E_{n\pm 1,\pm \mu}$) at certain magnetic fields. This degeneracy,
if occuring at the Fermi level, gives rise to a resonant
ISHC\cite{Shen}. The Hamiltonian for a 2DEG in a perpendicular
magnetic field is
\begin{eqnarray}
H_{e} &=&\frac{1}{2m'}(\vec{p}+\frac{e\vec{A}}{c})^{2}+\frac{\lambda }{\hbar }%
\hat{z}\cdot (\vec{p}+\frac{e\vec{A}}{c})\times \vec{\sigma}  \nonumber \\
&&-\frac{1}{2}g'_{s}\mu _{b}B\sigma _{z},
\end{eqnarray}%
where $m'$, $g'_{s}$ are the electron's effective mass and Lande-
$g$ factor, respectively. In addition, $\mu _{b}$ is the Bohr
magneton, $\lambda $ is the Rashba coupling. The eigen energy of
$H_{e}$ is given by
\begin{equation}
\epsilon _{ns}=\hbar \omega_{e} \left( n+\frac{\mu}{2}\sqrt{(1-g')^{2}+8n\eta ^{2}}%
\right) \,,
\end{equation}%
where $\omega_{e}=eB/m'c$ with $\mu=\pm 1$, for $n\geq 1$; and
$\mu=1$ for $n=0$. Numerically, the degeneracies between different
Landau levels ($E_{n\mu}$, $E_{n\pm 1,\pm \mu}$) request that
\begin{equation}
0<g'<1,
\end{equation}
for a reasonable value of Rashba coupling $\lambda/\hbar$. To a
turn, a set of parameters appropriate for
In$_{0.53}$Ga$_{0.47}$As/In $_{0.52}$Al$_{0.48}$As.\cite{Nitta97prl}
are in this range for $g'=0.1$.

Let's discuss the effect of Zeeman splitting on the ISHC of a 2DHG.
The 2D heavy-hole described by $k$-cubic Rashba model in a
perpendicular magnetic field including Zeeman splitting can be
written as
\begin{equation}
 H_{z}=\hbar\omega[(a^{\dag}a+\frac{1}{2})
+i\gamma(a^3\s^+-{a^{\dag}}^3\s^-)]+\frac{3}{2}g_{s}\mu _{b}B\sigma
_{z}
\end{equation}
and the energy levels in units of $\hbar\omega$ can be expressed as
\begin{equation}
E_{ns}=(n-1)+\frac{s}{2}\sqrt{(3-3g)^{2}+4\gamma ^{2}n(n-1)(n-2)},
\end{equation}
where $g=g_{s}m/2m_{e}$. The key difference between $k$-linear and
$k$-cubic Rashba model is the constant along with $g$, and ``3" in
$(3-3g)^{2}$ is due to $k$-cubic Rashba term, which is similar as
``1'' in $(1-g')^{2}$ arising from $k$-linear Rashba term. From
Eq.(12), we can deduce that if the energy crossing between $E_{ns}$
and $E_{n\pm 1,\mp s}$ occurs, which may give rise to a resonant
ISHC, requests that $\frac{2}{3}<g<1$ for a reasonable value of
$am/\hbar^{2}$. In calculation, we take the effective mass
$m=0.27m_{e}$(Ref.\cite{Wunderlich,Liu}), and then the corresponding
effective gyromagnetic factor $g_{s}$ is in the range of $4.93\sim
7.4$. A typical value of $g_{s}$ has been given by
Winkler\cite{Winkler}, and $g_{s}=-0.44$. The requested $g_{s}$ is
far away from the typical value of $g_{s}=-0.44$, which means that
the Zeeman splitting is too small to introduce ``effective"
degeneracy between different Landau levels. So the resonance effect
stems from energy crossing of different Landau levels near the Fermi
level due to the competition of Zeeman splitting and $k$-cubic
Rashba SOC is out of reach for real materials or in experiments.

In addition, the Zeeman splitting plays a crucial role on the ISHC
within $k$-linear Rashba model. When Zeeman energy is neglected, the
zero ISHC persists in a magnetic field\cite{Rashba}. Within the
perturbation method in this paper, we can also achieve this
conclusion. However, the Zeeman splitting has much less effect on
the ISHC of a 2DHG in magnetic field. To demonstrate this, a
resonant ISHC within Luttinger model including linear Zeeman
splitting term has been shown in Fig.3 (b), and parameter used are
taken from Ref.\cite{Bernevig,Ma} except $n_{h}=2.9\times
10^{16}$\,m$^{-2}$, and $g_{s}=-0.44$. The resonance appears at
$B=29.98T$. Therefore, the resonant ISHC of a 2DHG in a
perpendicular magnetic field contributed from the interplay between
mostly spin-$-\frac{1}{2}$ holes and spin-$\frac{3}{2}$ holes is
confirmed.

To compare with the previous result\cite{Loss} in the absence of
magnetic field, we calculate the interband contribution to the ISHC
as a function of the length scale $am/\hbar^{2}$ of the Rashba
coupling at a hole density of $n_{h}=3\times10^{14}m^{-2}$ for a
fixed magnetic field. Our numerical result has been shown in Fig.4.
The circle red dots indicates $\nu=100$ and the blue triangle
indicate $\nu=1000$. The ISHC in the absence of magnetic field as a
function of $am/\hbar^{2}$ at the same hole density has been plotted
in this fig too, which is denoted by the solid black line. To take a
overview expediently, it has also been shown in the fig inset solely
and data are obtained within the method in Ref.\cite{Loss}. As seen
in Fig.4, the interband contribution to the ISHC increases with
increasing Rashba coupling, and it approaches the value in absence
of magnetic field in weak magnetic limit.

\begin{figure}
\includegraphics[scale=0.45]{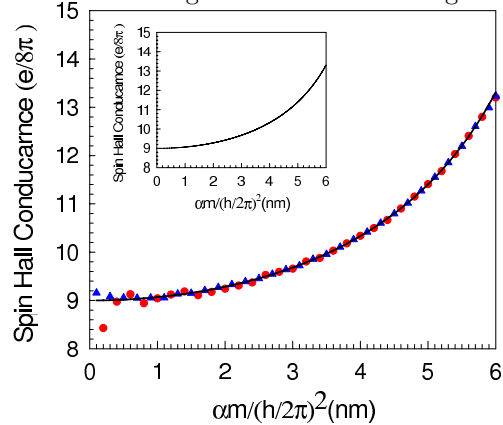}
\caption{(Color online) The interband contribution to the ISHC as a
function of $am/\hbar^{2}$ at $n_{h}=3\times10^{14}m^{-2}$ for a
fixed magnetic field. The circle red dots indicates $\nu=100$ and
the blue triangle indicates $\nu=1000$. Solid line and inset: The
ISHC as a function of $am/\hbar^{2}$ at the same hole density in the
absence of magnetic field, and data are achieved within the method
in Ref.\cite{Loss}.}\label{chargedensity}
\end{figure}

In conclusion, two likely ways to establish ISHE in experiments
shall be suggested. (1) A resonant ISHC shall appear in a 2DHG for a
high hole density and in a strong perpendicular magnetic field, and
this phenomenon may be used to distinguish the ISHE from the
extrinsic one. (2) The ISHC shows rich sign changes, and a notable
sign changes can be achieved for a lower hole density and moderate
magnetic field. The sign of ESHE depends on the sign of scattering
potential\cite{Yao,Fivaz}, and such difference may be used to
distinguish ESHE from intrinsic contributions and determine the
intrinsic origin of the effect. The sign issue should be regarded as
a very important aspect of SHE, and can be used in future
experiments or application.

We thank Shou-Cheng Zhang, Shun-Qing Shen and Wuming Liu for many
helpful discussions.


\begin{thebibliography}{99}
\bibitem{Prinz98Science} G. A. Prinz, Science 282, 1660 (1998); S. A. Wolf, D. D. Awschalom, R. A. Buhrman,
J. M. Daughton, S. von Moln¨¢r, M. L. Roukes, A. Y. Chtchelkanova,
and D. M. Treger, Science 294, 1488 (2001).
\bibitem{Review} D. Awschalom, D.Loss and N. Samarth, \emph{Semiconductor
Spintronics and Quantum Computation}, Springer 2002.
\bibitem{Murakami03Science} S. Murakami, N. Nagaosa, and S. C. Zhang,
Science \textbf{301}, 1348 (2003).
\bibitem{Sinova04} J. Sinova, D. Culcer, Q. Niu, N. A. Sinitsyn, T. Jungwirth,
and A. H. MacDonald, Phys. Rev. Lett. \textbf{92}, 126603 (2004).
\bibitem{Hirsch}J. E. Hirsch, Phys. Rev. Lett. \textbf{83}, 1834
(1999).
\bibitem{Wunderlich} J.
Wunderlich, B. Kaestner, J. Sinova, and T. Jungwirth, Phys. Rev.
Lett. {\bf 94}, 047204 (2005).
\bibitem{Bernevig} B. A. Bernevig and S. C. Zhang, Phys. Rev. Lett. {\bf 95}, 016801 (2005).
\bibitem{Kato}Y. K. Kato, R. C. Myers, A. C. Gossard, D. D.
Awschalom, Science {\bf 306}, 1910 (2004).
\bibitem{Engel} H. A. Engel, B. I. Halperin, and E. I. Rashba, Phys. Rev. Lett. {\bf 95}, 166605 (2005).
\bibitem{Bernevig0}B. A. Bernevig and S. C. Zhang, cond-mat/0412550.
\bibitem{Zarea}M. Zarea and S. E. Ulloa, Phys. Rev. B {\bf 73},
165306 (2006).
\bibitem{Ma}Tianxing Ma, Qiu Liu, Phys. Rev. B {\bf 73},
245315 (2006).
\bibitem{Sch}J. Schliemann, Int. J. Mod. Phys. B, {\bf 20}, 1015 (2006).
\bibitem{Shen}Shun-Qing Shen, M. Ma, X. C Xie, and F. C. Zhang,
Phys. Rev. Lett. {\bf 92}, 256603 (2004).
\bibitem{Dai}X. Dai, Z. Fang, Y. G. Yao, F. C Zhang, Phys. Rev. Lett. {\bf 96}, 086802
(2006).
\bibitem{Yao}Y. Yao, Z. Fang, Phys. Rev. Lett. {\bf 95}, 156601 (2005).
\bibitem{Shens}Shun-Qing Shen, Phys. Rev. B {\bf 70},
081311 (2004).
\bibitem{Loss}J. Schliemann and D. Loss, Phys. Rev. B {\bf 71},
085308 (2005).
\bibitem{Liu}S. Y. Liu and X.L. Lei, Phys. Rev. B {\bf 72}.
155314 (2005).
\bibitem{Winkler0} R. Winkler, H. Noh,
E. Tutuc, and M. Shayegan, Phys. Rev. B {\bf 65}, 155303 (2002).
\bibitem{Nitta97prl} J.Nitta, T. Akazaki, H. Takayanagi, T. Enoki, Phys. Rev. Lett.
78, 1335 (1997).
\bibitem{Winkler}R. Winkler, \emph{Spin-Orbit Coupling Effects in Two-Dimensional Electron and Hole System }(Springer, New York, 2003).
\bibitem{Rashba}E. I. Rashba, Phys. Rev. B \textbf{70}, 201309(R) (2004).
\bibitem{Fivaz}R. C. Fivaz, Phys. Rev. {\bf 183}, 586 (1969);
A. Cr\'{e}pieux, P. Bruno, Phys. Rev. {\bf B 64}, 014416 (2001).
\end{thebibliography}
\end{document}